 \documentclass[preprint,review,12pt]{elsarticle}

\pdfoutput=1

\usepackage{amssymb}

\usepackage{lineno}
\usepackage{algorithm}
\usepackage{algorithmic}
\usepackage{booktabs}
\usepackage{threeparttable} 
\usepackage{subfigure}
\usepackage{ulem}
\usepackage{epstopdf}

\usepackage{comment}

\usepackage{graphicx,amsmath,amsthm,amsfonts,xcolor,bm,url,fullpage}

\biboptions{square,comma,compress}

\newcommand*\patchAmsMathEnvironmentForLineno[1]{%
  \expandafter\let\csname old#1\expandafter\endcsname\csname #1\endcsname
  \expandafter\let\csname oldend#1\expandafter\endcsname\csname end#1\endcsname
  \renewenvironment{#1}%
     {\linenomath\csname old#1\endcsname}%
     {\csname oldend#1\endcsname\endlinenomath}}%
\newcommand*\patchBothAmsMathEnvironmentsForLineno[1]{%
  \patchAmsMathEnvironmentForLineno{#1}%
  \patchAmsMathEnvironmentForLineno{#1*}}%
\AtBeginDocument{%
\patchBothAmsMathEnvironmentsForLineno{equation}%
\patchBothAmsMathEnvironmentsForLineno{align}%
\patchBothAmsMathEnvironmentsForLineno{flalign}%
\patchBothAmsMathEnvironmentsForLineno{alignat}%
\patchBothAmsMathEnvironmentsForLineno{gather}%
\patchBothAmsMathEnvironmentsForLineno{multline}%
}

\newcommand{\tensor}[1]{\bm{\mathsf{#1}}}

\def\diff{\mathrm{d}}

\def\CFLmax{\mbox{CFL}_{\max}}

\def\Ma{\mbox{Ma}}

\def\eg{\textit{e.g.}}

\def\ie{\textit{i.e.}}

\journal{Computer $\&$ Fluids}

\begin{document}

\begin{frontmatter}

\title{Efficient $p$-multigrid method based on an exponential time discretization for compressible steady flows}


\author[csrc]{Shu-Jie Li\corref{mycorrespondingauthor}}
\cortext[mycorrespondingauthor]{Corresponding author}
\ead{shujie@csrc.ac.cn}
%
%
%
%

\address[csrc]{Beijing Computational Science Research Center,  Beijing 100193,
  China}
 %

%
%

\begin{abstract}
An efficient multigrid framework is developed for the time marching of steady-state compressible flows with a spatially high-order ($p$-order polynomial) modal discontinuous Galerkin method. 
The core algorithm that based on a global coupling, exponential time integration scheme provides strong damping effects to accelerate the convergence towards the steady state, while high-frequency, high-order spatial error modes are smoothed out with a $s$-stage  preconditioned Runge-Kutta method.
 Numerical studies show that the exponential time integration substantially improves the damping and propagative efficiency of 
 Runge-Kutta time-stepping for use with the $p$-multigrid method,
 yielding rapid and  $p$-independent convergences to steady flows in both two and three dimensions.
 \end{abstract}


\end{frontmatter}


\section{Introduction}
\label{sec:intro}

An important requirement for computational fluid dynamics (CFD) applications  is the capability to predict steady flows such as the case
 of flow past a complex geometry, so that key performance parameters {\textit {e.g.}}, the lift and drag coefficients  can be estimated.
While the classical second-order methods are still being used extensively, high-order spatial discretizations attract increasing interests.
However, for steady-state computations, most of the spatial discretizations have rested on the use of limited, traditional time discretizations
 combining with various acceleration methods. Among these methods, the $p$-multigrid acceleration is natural in the context of modal
 discontinuous Galerkin methods (DG) where the accuracy variation can be realized by truncating the DG polynomial.
For solving steady-state problems, the performance of a $p$-multigrid process depends on the efficiency of time advancement method or say smoother used. 
Unfortunately, in contrast to a relative ubiquity of spatial discretizations, efficient time-marching approaches seem to be limited.  
Recently, as an alternative to traditional time-marching methods, 
an exponential time integration scheme,  predictor-corrector exponential time-integrator scheme (PCEXP) \cite{Shujie17AIAA,Shujie18AIAA,Shujie18JCP}, is developed and successfully applied to the time stepping of CFD  problems, exhibiting some advantages in terms of accuracy and efficiency for solving the fluid dynamics problems governing by the Euler and Navier-Stokes equations for either time-dependent and time-independent regimes.

In this paper, the exponential time integration is exploited in a new $p$-multigrid framework that consisting of an exponential time marching method and a $s$-stage preconditioned Runge-Kutta method as an effective way to increase the feasibility of arbitrarily $p$-order DG for the high-order simulations of steady-state flows. The framework combines the good numerical damping feature of the exponential scheme and the low-memory feature of a preconditioned Runge-Kutta (PRK) method, resulting in low-cost 
and memory-friendly features for high-order computations of steady flows.

The remainder of this paper is organized as follows.
Section~\ref{sec:pmultigrid}  presents the multigrid algorithm which combines two stand-alone methods in a V-cycle $p$-multigrid framework.
Section~\ref{sec:space} introduces the spatial discretization with a modal high-order DG method.
Section~\ref{sec:timestepping} discusses how to evaluate the time steps in the $p$-multigrid framework.
Section~\ref{sec:results} presents the numerical results 
including two inviscid flow problems at Mach number $\Ma=0.3$: (a) flow past a circular cylinder; (b) flow flow over a sphere.
The numerical results obtained with the exponential $p$-multigrid method (eMG)  are compared directly with a fully implicit method solved with the Incomplete LU preconditioned GMRES (ILU-GMRES) linear solver.
Finally, Section~\ref{sec:final} concludes this work.
The Appendix provides the details of Jacobian matrices for the DG space discretization and time-step evaluations.

\section{The $p$-multigrid frame}
\label{sec:pmultigrid}
Although $p$-multigrid methods are proved effective for high-order computations of steady flows, the iteration count and the computational cost vary greatly.
Actually, $p$-multigrid methods are hard to be compared to fully implicit methods. In this section, a new $p$-multigrid frame is detailed which is expected to have comparable performance to implicit methods for steady-state computations. The algorithm combines two stand-alone methods: the \textit {exponential time integration} method and a $s$-stage \textit {preconditioned Runge-Kutta} method.
The two methods are introduced separately first and finally shown to be integrated into a whole V-cycle multigrid frame.

\subsection{Exponential time integration}
\label{sec:exp}
The first-order exponential time integration scheme (EXP1) is presented 
and followed by the details of an efficient implementation through the Krylov method. 

We start with the following semi-discrete system of autonomous
ordinary differential equations which may be obtained from a spatial
discretization:
\begin{equation}
  \frac{\diff \mathbf{u}}{\diff t} 
  = 
  \mathbf{R}(\mathbf{u}) 
  ,
\label{eqn:starteq}
\end{equation}
where $\mathbf{u} = \mathbf{u}(t) \in \mathbb{R}^K$ denotes the vector
of the solution variables and $\mathbf{R}(\mathbf{u}) \in
\mathbb{R}^K$ the right-hand-side term which may be the spatially
discretized residual terms of the discontinuous Galerkin method used
in this work.  The dimension $K$ is the degrees of freedom which can
be very large for 3-D problems.
Without loss of generality, we consider $\mathbf{u}(t)$ in the
interval of one time step, \ie, $t \in [t_n,\, t_{n+1}]$.
Applying  the linearization splitting method \cite{Caliari2009anm} to Eq.~\eqref{eqn:starteq}
 leads to a different exact expression 
\begin{equation}
  \frac{\diff \mathbf{u}}{\diff t} 
  =
  \tensor{J}_{n} \mathbf{u}
  + \mathbf{N}(\mathbf{u})
  ,
\label{eqn:starteq2}
\end{equation}
where the subscript  $n$ indicates the value evaluated at $t =
t_n$, $\tensor{J}_{n}$ denotes the Jacobian matrix 
$
\tensor{J}_{n} 
=
\partial \mathbf{R}(\mathbf{u}) / \partial \mathbf{u} |_{t = t_n}
= \partial \mathbf{R}(\mathbf{u}_n) / \partial \mathbf{u}
$
and
$
\mathbf{N}(\mathbf{u}) 
= 
\mathbf{R}(\mathbf{u}) 
- \tensor{J}_{n}{\mathbf{u}}
$
denotes the remainder, which in general is nonlinear.
Eq.~\eqref{eqn:starteq2} admits the following formal solution:
\begin{equation}
  \mathbf{u}_{n+1}  
  = 
  \exp (\Delta t \tensor{J}_n)  \mathbf{u}_n 
  +
  \int_{0}^{\Delta t} 
  \exp \left( (\Delta t-\tau) \tensor{J}_n \right)
  \mathbf{N} ( \mathbf{u}(t_n + \tau) )
  \,
  \diff \tau 
  ,
\label{eqn:s3}
\end{equation}
where $\Delta t = t_{n+1}-t_n$ and
\begin{equation}
  \exp(- t \tensor{J}_n) 
  = 
  \sum_{m=0}^\infty  
  \frac{ \left(- t \tensor{J}_n \right)^m}{m!}
\end{equation}
is the integrating factor.
The formal solution \eqref{eqn:s3} is the starting point to derive the
proposed exponential scheme in which the stiff part is computed
analytically whereas the nonlinear term is approximated numerically.
By approximating the nonlinear term $\mathbf{N}$ by a constant, \ie, its left-end value $\mathbf{N}_n$ on the interval $[t_n,\, t_{n+1}]$, hence,
\begin{equation}
  \mathbf{N}(\mathbf{u}(t_n+\tau))
  \approx 
  \mathbf{N}_n
  =
  \mathbf{R}_n 
  - 
  \tensor{J}_n 
  \mathbf{u}_n 
  , 
\end{equation}
leading to the first-order exponential scheme EXP1:
\begin{equation}
  \mathbf{u}_{n+1}
  =
  \exp (\Delta t \tensor{J}_n) 
  \, \mathbf{u}_n + {\Delta t} \, \tensor{\Phi}_1(\Delta t \tensor{J}_n) 
  \, \mathbf{N}_n 
  =
  \mathbf{u}_n + {\Delta t} \, \tensor{\Phi}_1(\Delta t \tensor{J}_n) 
  \, \mathbf{R}_n 
  ,
\label{eqn:EXP1}
\end{equation}
where
\begin{equation}
  \tensor{\Phi}_1( \Delta t \tensor{J} )
  :=
  \frac{\tensor{J}^{-1}}{\Delta t}
  \left[
  \exp\left( \Delta t \tensor{J} \right)
  - \tensor{I}
  \right]
  ,
\label{eqn:Phi1}
\end{equation}
and $\tensor{I}$ denotes the $K \times K$ identity matrix.

The physical nature of such type of exponential schemes
relies on the global coupling feature via the global Jacobian matrix $\tensor J$, 
so that flow transportation information can be broadcasted to the whole computational domain without a \mbox{CFL} restriction.  
That is why the exponential schemes behavior like a fully implicit method but only depends on the current solution, 
 \ie, in an explicit way as Eq.~\eqref{eqn:EXP1}. While the second-order PCEXP scheme is more appropriate for computing unsteady problems, 
the EXP1 scheme which is the first-order PCEXP scheme is shown to be especially effective for steady flows as
in reference \cite{Shujie17AIAA, Shujie18JCP}. 
So in this paper, the EXP1 scheme is  curiously exploited in the $p$-multigrid framework for steady flow computations.

\subsection{Realization of EXP1 with the Krylov method}

The implementation of exponential time integration schemes requires evaluations of
matrix-vector products, and in particular, the product of the
exponential functions of the Jacobian and a vector, \eg,
$\tensor{\Phi}_1 ( \Delta t \tensor{J}_n ) \mathbf{N}$ in \eqref{eqn:EXP1}.
If the inverse of the Jacobian $\tensor{J}$ exists, then it is
possible to use $\tensor{J}^{-1}$ to compute $\tensor{\Phi}_1 (\Delta
t \tensor{J})$ defined by \eqref{eqn:Phi1}.
However, $\tensor{J}$ may be singular and hard to be computed,
 \eg, in the presence of periodic boundary conditions. 
 In addition, for a problem with a very large number of
degrees of freedom, direct inversion of $\tensor{J}$ can be
prohibitively expensive to compute. 
Instead, the matrix-vector products can be approximated
efficiently using the Krylov method \cite{Tokman2010pcs,
  Saad1992siamjna}, which can also treat a singular $\tensor{J}$.  The
 basic idea of the Krylov method is to approximate the product of
$\exp(\Delta t \tensor{J})$ and a vector, such as $\mathbf{N}$,
 by projecting it onto a small Krylov subspace, resulting
in a much smaller matrix thus cheaper in computational cost.  

We start the algorithm with the Taylor expansion of $\exp(\Delta t \tensor{J})$, 
and the product $\tensor{\Phi}_1 \mathbf{N}$ can be written as:
\begin{equation}
  \tensor{J}^{-1} 
  \frac{\exp(\Delta t \tensor{J}) - \tensor{I}}{\Delta t} \mathbf{N} 
  = 
  \sum_{k=0}^\infty
  \frac{ (\Delta t \tensor{J})^k}{(k+1)!} \mathbf{N}
  =
  \left( 
  \tensor{I} 
  + \frac{(\Delta t \tensor{J})}{2!} 
  + \frac {(\Delta t \tensor{J})^2}{3!} 
  + \cdots 
  \right)  
  \mathbf{N}
  .
\end{equation}
It can be approximated by the following function projection onto the
Krylov subspace of dimension $m$:
\begin{equation}
  \mathbb{K}_m(\tensor{J}, \mathbf{N}) 
  = 
  \mbox{span} 
  \{ \mathbf{N},\, 
  \tensor{J} \mathbf{N},\, 
  \tensor{J}^2 \mathbf{N},\, 
  \ldots,\, 
  \tensor{J}^{m-1} \mathbf{N} \}
  .
\end{equation}
The orthogonal basis matrix $\tensor{V}_m := ( \mathbf{v}_1,\,
\mathbf{v}_2,\, \cdots,\, \mathbf{v}_m ) \in \mathbb{R}^{K \times m}$
satisfies the so-called Arnoldi decomposition \cite{Saad1992siamjna}:
\begin{equation}
  \tensor{J} \tensor{V}_m 
  =
  \tensor{V}_{m+1}  \widetilde{\tensor{H}}_m,
\label{eqn:vm}
\end{equation}
where $\tensor{V}_{m+1} := ( \mathbf{v}_1,\, \mathbf{v}_2,\, \cdots,\,
\mathbf{v}_m,\, \mathbf{v}_{m+1} ) = ( \tensor{V}_{m},\,
\mathbf{v}_{m+1} ) \in \mathbb{R}^{K \times (m+1)}$. 
%
%
The $(m+1) \times m$ upper-Hessenberg matrix $\widetilde{\tensor{H}}_m$ can be written
as 
\begin{equation}
  \widetilde{\tensor{H}}_m 
  =
  \begin{bmatrix}
    \tensor{H}_{m}
  \\
  h_{m+1,m} \mathbf{e}_m^{\tensor{T}} 
  \end{bmatrix}
  ,
\end{equation}
where $\tensor{H}_{m}$ is the matrix composed of the first $m$ rows of
$\widetilde{\tensor{H}}_m$ and $\mathbf{e}_m := (0,\, \cdots,\, 0,\,
1)^{\tensor{T}} \in \mathbb{R}^m$ is the $m$-th canonical basis vector
in $\mathbb{R}^m$, then Eq.~\eqref{eqn:vm} becomes
\begin{equation}
  \tensor{J} \tensor{V}_m
  =
  \tensor{V}_m \tensor{H}_m 
  + h_{m+1,m} \mathbf{v}_{m+1} \mathbf{e}_m^{\tensor{T}} 
  . 
\label{vm1}
\end{equation}
Because $\tensor{V}_m^{\tensor{T}} \tensor{V}_m = \tensor{I}$,
therefore
\begin{equation}
  \tensor{H}_m 
  =
  \tensor{V}^{\tensor{T}}_m \tensor{J} \tensor{V}_m 
  ,
\label{vm2}
\end{equation}
that is, $\tensor{H}_m$ is the projection of the linear transformation
of $\tensor{J}$ onto the subspace $\mathbb{K}_m$ with the basis
$\mathbb{V}_m$.
Since $\tensor{V}_m \tensor{V}_m^{\tensor{T}} \neq \tensor{I}$,
Eq.~\eqref{vm2} leads to the following approximation:
\begin{equation}
  \tensor{J}
  \approx
  \tensor{V}_m \tensor{V}^{\tensor{T}}_m 
  \tensor{J} 
  \tensor{V}_m \tensor{V}^{\tensor{T}}_m
  =
  \tensor{V}_m \tensor{H}_m \tensor{V}^{\tensor{T}}_m 
  ,
\end{equation}
and $\exp (\tensor{J})$ can be approximated by $\exp ( \tensor{V}_m
\tensor{H}_m \tensor{V}^{\tensor{T}}_m )$ as below
\begin{equation}
  \exp (\tensor{J}) \mathbf{N}   
  \approx   
  \exp ( \tensor{V}_m \tensor{H}_m  \tensor{V}^{\tensor{T}}_m)  
  \mathbf{N} 
  = 
  \tensor{V}_m  
  \exp  ( \tensor{H}_m )  \tensor{V}^{\tensor{T}}_m   \mathbf{N}
  .
\label{evv}
\end{equation}
The first column vector of $\tensor{V}_m$ is $\mathbf{v}_1 =
\mathbf{N}/ \| \mathbf{N} \|_2$ and $\tensor{V}^{\tensor{T}}_m
\mathbf{N} = \| \mathbf{N} \|_2 \, \mathbf{e}_1$,
 thus \eqref{evv} becomes
\begin{equation}
  \exp (\tensor{J}) \mathbf{N} 
  \approx   
  \| \mathbf{N} \|_2  
  \tensor{V}_m \exp(\tensor{H}_m) \, {\mathbf{e}_1} 
  .
\end{equation}
Consequently, $\tensor{\Phi}_1$ can be approximated by
\begin{equation}
  \tensor{\Phi}_1(\Delta t  \tensor{J})  \mathbf{N} 
  = 
  \frac {1}{{\Delta t}}
  \int_{0}^{\Delta t} \hspace*{-2ex} \exp ((\Delta t-\tau)
  \tensor{J})   
  \mathbf{N} \, 
  \diff \tau 
  \approx 
  \frac {1}{{\Delta t}}  
  \int_{0}^{\Delta t} \hspace*{-2ex}  \| \mathbf{N} \|_2  \tensor{V}_m  
  \exp \left((\Delta t-\tau) \tensor{H}_m \right) 
  \, {\mathbf{e}_1} \, \diff \tau.
\label{final1}
\end{equation}
In general, the dimension of the Krylov subspace, $m$, is chosen to be
much smaller than the dimension of $\tensor{J}$, $K$, thus
$\tensor{H}_m \in \mathbb{R}^{m\times m}$ can be inverted easily,
so $\tensor{\Phi}_1$ can be easily computed as the following
\begin{align}
  \tensor{\Phi}_1(\Delta t \tensor{J})  \mathbf{N}  
  &\approx 
  \frac {1}{\Delta t}  
  \| \mathbf{N} \|_2 \tensor{V}_m  
  \int_{0}^{\Delta t}
  \exp \left((\Delta t-\tau) \tensor{H}_m \right) \, 
  \mathbf{e}_1 \, \diff \tau
  \nonumber 
  \\
  &= 
  \frac {1}{{\Delta t}}  \|\mathbf{N}\|_2  
  \tensor{V}_m   \tensor{H}_m^{-1}  
  \left[ \exp (\Delta{t}   \tensor{H}_m) - \tensor{I} \right] 
  \mathbf{e}_1
  ,
\label{final}
\end{align}
where the matrix-exponential $\exp(\Delta t \tensor{H}_m)$ can be
computed efficiently by the Chebyshev rational approximation (cf.,
\eg, \cite{Saad1992siamjna, Moler2003siamjna}) due to the small size
of $\tensor{H}_m$.

\subsection{Preconditioned Runge-Kutta method}

Consider a $s$-stage preconditioned Runge-Kutta (PRK) method of the following form
\begin{align}
{\bf u}^{(0)} &= {\bf u}^n \nonumber \\
{\bf u}^{(k)} &={\bf u}^{n} + \beta_k \, \mathbf{P}^{-1}({\bf u}^n) \, \mathbf R \left ( {\bf u}^{(k-1)} \right ),\quad k=1,2,\dots,s \label{prk_scalar} \\
{\bf u}^{n+1} &={\bf u}^{(s)}  \nonumber 
\end{align}
where $\beta_k = 1/(s-k+1)$. $\bf P$ is taken as the diagonal part of the global residual Jacobian ${\tensor J}=\partial{\bf R}/{\partial{\bf u}}$, 
representing the element-wise wave propagation information. $s=4$ is used for all the test cases of this work.

The physical nature of this type of RK method can be interpreted in two different views which are helpful for us to see how does PRK make sense.
First, we consider the first-order spatial discretization of finite volume or discontinuous Galerkin method to the $i$-th element surrounded
 by adjoined  cells  $j$ ($1 \le j \le N$)  with the inter-cell surface area  ${\rm S}_{ij}$, and the spatial residual using a upwinding flux can be  written as
\begin{equation}
V_i \frac{\Delta {\bf u}_i}{\Delta t}={\bf R}_i = \sum_{j=1}^{N}{\frac{1}{2} \left[{\bf{F}} ({\bf u}_i) + {\bf{F}} ({\bf u}_j) \right]{\bf n}_{ij} {\rm S}_{ij} + 
\frac{1}{2}|{\bf A}_{ij}^{n}| \left( {\bf u}_i - {\bf u}_j \right) {\rm S}_{ij} }
\end{equation}
So $\bf P$ can be derived as
\begin{equation}
{\bf P}_i = \frac{\partial{\bf R}_i}{\partial{\bf u}_i}  
\approx \sum_{j=1}^{N}{ \frac{1}{2}|{\bf A}_{ij}^{n}|{\rm S}_{ij} }
\end{equation}
A matrix ${\bm {\Delta  t}}$  can be defined as
\begin{equation}
{\bm {\Delta  t}} = {V_i} {\bf P}^{-1}= \frac{V_i}{\sum_{j=1}^{N}{ \frac{1}{2}|{\bf A}_{ij}^{n}|{\rm S}_{ij} }} 
\end{equation}
One can uncover the relationship between the matrix ${\bm {\Delta  t}}$  and the traditional definition of time step 
by considering a cell-constant scalar spectral radius approximation $\lambda_{ij}^{max}$ to $|{\bf A}_{ij}^n|$ , \ie,
\begin{equation}
\Delta {t} = \frac{2 V_i}{\sum_{j=1}^{N}{{\lambda_{ij}^{max}}{\rm S}_{ij} }} \overset{1D} \longrightarrow \frac{\Delta x_i}{\lambda_{i}^{max}}
\end{equation}
Therefore, ${\bf P}^{-1}$ is equivalent to a matrix time step and it is consistent to the usual definition of time step in the scalar case.
Nile \cite{pierce1996preconditioning} demonstrated that this matrix is a kind of preconditioner which can provide effective clustering of convective eigenvalues 
and substantial improvements to the convergence of RK time-stepping. In this work, different from Nile's approximation to ${\bf P}^{-1}$,
an exact  way of evaluating matrix time steps with exact Jacobian is proposed in Section 4.3,
 so that all the stiffness effects from spatial discretizations and boundary conditions can be exactly taken into account.

In the second view, the PRK scheme is found to be a simplified implicit method without considering the contributions from off-diagonal terms.  
To see how does it make sense, we consider a standard implicit discretization to a steady equation
\begin{equation}
{\bf R}({\bf u}_{n+1}) = {\bf R}({\bf u}_{n}) + \tensor J \, ({\bf u}_{n+1}-{\bf u}_{n})=0
\end{equation}
where $\tensor J=\tensor D+\tensor O$, $\tensor D$ and $\tensor O$ denotes the diagonal and off-diagonal parts of $\tensor J$, respectively. If we ignore the contribution from the off-diagonal part, namely  $\tensor J\approx \tensor D$, we have
\begin{equation}
{\bf u}_{n+1} = {\bf u}_{n} - {\tensor J}^{-1} {\bf R}({\bf u}_{n}) \approx
{\bf u}_{n} - {\tensor D}^{-1}\, {\bf R}({\bf u}_{n})
\end{equation}
So the PRK scheme is actually a point implicit scheme which dismiss the inter-cell coupling and can be considered as a multi-stage block Jacobi iteration. By dismissing the off-diagonal terms, neighboring cells are decoupled, so stability issues might occur when using a large CFL number.
To cure this problem, we recommend to use ${\tensor J}=\partial{\bf R}/{\partial{\bf u}} + {\tensor I}/\delta \tau$. $\delta \tau$ is used for increasing diagonal domination which is computed as \eqref{x3d} with $\mbox{CFL}=100$ for all test cases, resulting a cheap way for the multigrid smoothing of high frequency errors.

\subsection{The V-cycle $p$-multigrid framework }
The use of $p$-multigrid smoother with explicit RK or preconditioned RK methods is observed inefficient at eliminating low-frequency error modes at lower orders of accuracy. 
To provide a better smoother with stronger damping effects,
the EXP1 scheme that exhibits fast convergence rates for Euler and Navier-Stokes equations is considered.  Unlike the explicit RK smoother that only produces weak damping effects in a local, point-wise manner, the exponential scheme is a global method that allows large time steps with strong damping effects to all the frequency modes across the computational domain, as shown in the previous works \cite{Shujie18JCP}.

In the exponential $p$-multigrid method (eMG),  the EXP1 scheme is utilized on the accuracy level $p=0$ and the PRK method is used for accuracy levels  $p>0$, 
contributing both memory deduction and efficiency enhancement.
The smoothing employs a V-cycle $p$-multigrid process, where a two-level algorithm is recursively used. To illustrate the algorithm, let us consider a nonlinear problem ${\bf A}({\bf u}^p)={\bf p}^p$, where ${\bf u}^p$ is the solution vector, ${\bf A}({\bf u}^p)$ is the nonlinear operator and $p$ denotes the accuracy level $p$.
Let ${\bf v}^p$ be an approximation to the solution vector ${\bf u}^p$ and define the residual ${\bf r}({\bf v}^p)$ by
\begin{equation}
{\bf r}({\bf v}^p) = {\bf f}^p - {\bf A}^p({\bf v}^p) \nonumber
\end{equation}
In the eMG framework, the solution on the $p-1$ level is used to correct the solution of $p$ level in the following steps:
\begin{enumerate}
\item Conduct a time stepping with the PRK scheme on the highest accuracy level $p_{\scriptsize{\mbox{max}}}$.
\item Restrict the solution and the residual of $p$  to the $p-1$ level ($1 \le p \le p_{\scriptsize{\mbox{max}}}$)
\begin{equation}
{\bf v}^{p-1}_0 = \mathbb{R}_p^{p-1} {\bf v}^p, \quad
{\bf r}^{p-1} = \mathbb{R}_p^{p-1} {\bf r}^p({\bf v}^p)
\end{equation}
where $\mathbb{R}_p^{p-1}$ is the restriction operator from the level $p$ to the level $p-1$. 
\item Compute the forcing term for the $p-1$ level
\begin{equation}
{\bf s}^{p-1} = {\bf A}^{p-1}({\bf v}^{p-1}_0) - {\bf r}^{p-1}.
\end{equation}
\item Smooth the solution with the PRK scheme on the  $p-1$ level but switch to use the EXP1 scheme on the lowest accuracy level $p=0$,
\begin{equation}
{\bf A}^{p-1}({\bf v}^{p-1}) = \mathbb{R}_p^{p-1} {\bf f}^p + {\bf s}^{p-1}.
\end{equation}
\item Evaluate the error of level $p-1$
\begin{equation}
{\bf e}^{p-1} = {\bf v}^{p-1} - {\bf v}^{p-1}_0.
\end{equation}
\item Prolongate the $p-1$ error and correct the approximation of level $p$
\begin{equation}
{\bf v}^p = {\bf v}^p + \mathbb{P}_{p-1}^p {\bf e}^{p-1}
\end{equation}
where $\mathbb{P}_{p-1}^p$ is the prolongation operator.
\end{enumerate}

\section{Spatial discretization}
\label{sec:space}
In this paper, the eMG method is applied to solve three-dimensional Euler equations discretized
 by a modal discontinuous Galerkin method. Consider the Euler equations in a rotating frame of reference in three-dimensional space
\begin{equation}
  \frac{\partial \mathbf{U}}{\partial t} 
  + \bm{\nabla} \cdot \tensor{F} 
  = 
  \mathbf{S},
\label{euler}
\end{equation}
where $\mathbf{U}$ stands for the vector of
conservative variables, $\tensor{F}$ denotes
the convective flux, and $\mathbf{S}$ is the source
term
\begin{equation}
  \mathbf{U}
  = 
  \left( 
  \begin{array}{c} 
    \rho \\ 
    \rho \bm{v} \\
    \rho E
  \end{array} 
  \right) 
  ,
  \quad 
  \tensor{F}
  =
  \left(
  \begin{array}{c}
    \rho\, (\bm{v}-\bm{v}_r)^{\tensor{T}} 
    \\ 
    \rho\, (\bm{v}-\bm{v}_r)
    \bm{v}^{\tensor{T}} 
    +
    p \, \tensor{I}
    \\ 
    \rho\, H\, (\bm{v}-\bm{v}_r)^{\tensor{T}} 
  \end{array}
  \right) 
   ,
   \quad
  \mathbf{S}
  = 
  \left( 
  \begin{array}{c} 
    0 
    \\ 
    - \rho \, \bm{\omega} \times \bm{v} 
    \\
    0 
  \end{array} 
  \right) ,
\end{equation} 
where $\bm{v} = (u,\, v,\, w)^{\tensor{T}}$ is the absolute velocity,
$\bm{\omega} = (\omega_x,\, \omega_y,\, \omega_z)^{\tensor{T}}$ is
the angular velocity of the rotating frame of reference, $\bm{v}_r =
\bm{\omega} \times \bm{x}$; $\rho$, $p$, and $e$ denote the flow
density, pressure, and the specific internal energy; $E = e +
\frac{1}{2}||\bm{v}||^2$ and $H = E + p/\rho$ denote the total energy
and total enthalpy, respectively; $\tensor{I}$ denotes the $3 \times
3$ unit matrix; and the pressure $p$ is given by the equation of state for a perfect
gas
\begin{equation}
  p=\rho \left(\gamma-1\right) e,
\end{equation} 
where $\gamma=7/5$ is the ratio of specific heats for perfect
gas.

\subsection{Modal discontinuous Galerkin method}
\label{sec:DG}

Considering a computational domain $\Omega$ divided into a set of
non-overlapping elements of arbitrary shape,
the modal discontinuous Galerkin method seeks an approximation $\mathbf{U}_h$ in each element $E \in \Omega$
with finite dimensional space of polynomial $P^{p}$ of
order {$p$} in the discontinuous finite element space
\begin{equation}
  \mathbb{V}_h 
  :=
  \{ \psi_i \in L^{2}(\Omega): 
  \psi_i |_E \in {P}^{{\color{purple} p}}(\Omega),\, 
  \forall \, E \in \Omega \}
  .
\end{equation}
The numerical solution of $\mathbf{U}_h$ can be approximated in the
finite element space $\mathbb{V}_h$
\begin{equation}
  \mathbf{U}_h(\bm{x},\, t )
  =  
  \sum_{j = 1}^n 
  \mathbf{u}_j(t) \psi_j( \bm{x} )
  .
\label{uexp}
\end{equation}
In the weak formulation, the Euler equations \eqref{euler} in an
element $E$ becomes:
\begin{equation}
  \int_{E} 
  \psi_i \psi_j \diff \bm{x}
  \frac{\diff \mathbf{u}_j}{\diff t}  
  = 
  - \int_{\partial E} \psi_i \widetilde{\tensor{F}} 
  \cdot \hat{\bm{n}}
  \, \diff \bm\sigma
  +
  \int_{E} 
  ( \tensor{F} \cdot \bm{\nabla} \psi_i  + \psi_i  \mathbf{S} ) 
  \diff \bm{x}
  :=
  \mathbf{R}_i
  ,
\label{euler3}
\end{equation}
where $\hat{\bm{n}}$ is the out-normal unit vector of the surface
element $\bm \sigma$ with respect to the element $E$,
$\widetilde{\tensor{F}}$ is the Riemann flux \cite{Toro1999},
which will be approximated by Roe's flux \cite{Roe1981jcp},
and the Einstein summation convention is used.
For an orthonormal basis $\{ \psi_i \}$, the term on the left-hand
side of Eq.~\eqref{euler3} becomes diagonal, so the system is in the
standard ODE form of Eq.~\eqref{eqn:starteq}, thus avoiding solving a
linear system as required for a non-orthogonal basis.
More importantly, the use of orthogonal basis would yield more
accurate solutions, especially for high-order methods with $p \gg 2$.

\subsection{Orthonormal basis functions with the Cartesian coordinates}

In this paper, the basis function $\psi _i(\bm{x})$ is defined on the
global Cartesian coordinate $\bm{x}:=(x,\, y,\, z)$ rather than on the
cell-wise, local reference coordinates.  The variable values on the
Gaussian quadrature points for computing the surface
fluxes can be easily accessed without the Jacobian mapping between the
local reference coordinates to the global Cartesian ones
\cite{DengS2006anacm, Bergot2013nmpde}, and it also makes the
discontinuous Galerkin method feasible on arbitrary polyhedral grids
\cite{Botti2012siamjsc}.

A simple choice of the basis function in \eqref{uexp} may be the
monomials \cite{Wolkov2011} or Taylor basis \cite{LuoH2012aiaa0461}.
However, in the case of distorted meshes, the non-orthogonality of
these basis functions may yield an ill-conditioned mass matrix,
resulting in degradation of accuracy and even loss of numerical
stability.  In this work, to construct an orthonormal basis set $\{
\psi_i(\bm{x}) \}$, we start with the normalized monomials $\{
\chi_i(\bm{x}) \}$:
\begin{subequations}
\begin{align}
  {\{\chi_i(\bm{x})\}}
  &
  := 
  \left\{
  \left.
  \frac{ ( x - x_c )^{p_1} ( y - y_c)^{p_2} ( z - z_c)^{p_3}}
       {{L_x}^{p_1} {L_y}^{p_2} {L_z}^{p_3}}
        \right| 0 \le p_1,\, p_2,\, p_3 \, ;
       \ p_1 + p_2 + p_3 \le i-1 
  \right\}
  ,
\label{basis2}
\\
  {\{\psi_i(\bm{x}) \}}
  &
  :=  
  \left\{
  s_{i} 
  \left[
  \left.
    \chi_i(\bm{x}) 
    +
    \sum_{j = 1}^{i - 1} 
    c_{ij} \chi_j(\bm{x}) 
    \right]  \right|
  1 \le i \le N \right\},
\label{basis}
\end{align}
\end{subequations}
where 
\begin{align*}
  &
  x_c 
  := 
  \frac{1}{| E |} \int_{E} x \, \diff \bm{x} 
  , 
  &&
  y_c 
  := 
  \frac{1}{| E |} \int_E y \, \diff \bm{x} 
  , 
  &&
  z_c 
  := 
  \frac{1}{| E |} \int_E z \, \diff \bm{x} 
  , 
  \\
  &
  L_x 
  :=
  \frac{1}{2}\left(x_{\max} - x_{\min}\right)
  ,
  &&
  L_y
  :=
  \frac{1}{2}\left(y_{\max} - y_{\min}\right)
  ,
  &&
  L_z
  :=
  \frac{1}{2}\left(z_{\max} - z_{\min}\right)
  ,
\end{align*}
and the total number of basis functions $N = (p + 1)(p + 2)(p + 3)/6$
for the $p$-th order DG approximation in 3-D space.  With the following
definition of the inner product on an element $E$:
$
\left\langle f,\, g \right\rangle_E 
:= \int_E f \left(\bm{x}\right) g\left(\bm{x}\right) \, \diff \bm{x}
,
$
the coefficients $\{ s_{i} \}$ and $\{c_{ij}\}$ can be computed with
the modified Gram-Schmidt (MGS) orthogonalization described in \cite{Shujie18JCP}.

\subsection{Exact Jacobian matrix for the eMG method}

The broadcasting of global information is achieved through the exact Jacobian matrix which accurately includes
the information of both the interior and the boundary of the
elements, contributing enhancements of convergence rate and stability to the PCEXP and EXP1 schemes \cite{Shujie18JCP}, and in this paper,
it is also used for the evaluation of the matrix time steps in the PRK scheme.
The details of computing the exact Jacobian is outlined as follows.

The diagonal Jacobians are computed by taking the derivative of
\eqref{euler3} with respect to the $\mathbf{u}_j$ of the host cell
with the label ``L'':
\begin{align}
  \frac{\partial \mathbf{R}_i}{\partial \mathbf{u}_j^{\rm L} } 
  & 
  = 
  - \int_{\partial E} \psi_i 
  \frac{\partial \widetilde{\tensor{F}} (\mathbf{U}_{\rm L},\,
    \mathbf{U}_{\rm R} )} 
       {\partial \mathbf{U}_{\rm L}} 
  \frac{\partial \mathbf{U}_{\rm L}} 
       {\partial \mathbf{u}_j^{\rm L}}   
  \, \diff \bm{\sigma} 
  + 
  \int_{E} 
  \left( 
  \bm{\nabla} \psi_i   
  \frac{\partial \tensor{F}}{\partial \mathbf{U}} 
  \frac{\partial \mathbf{U}}{\partial \mathbf{u}_j} 
  + 
  \psi_i  
  \frac{\partial \mathbf{S}}{\partial \mathbf{U}} 
  \frac{\partial \mathbf{U}}{\partial \mathbf{u}_j}
  \right) 
  \diff \bm{x} 
  \nonumber
  \\
  & 
  = 
  - \int_{\partial E} 
  \psi_i^{\rm L} \, 
  \psi_j^{\rm L} \, 
  \frac{\partial \widetilde{\tensor{F}}(\mathbf{U}_{\rm L},\,
    \mathbf{U}_{\rm R}) }
       {\partial \mathbf{U}_{\rm L}}  
  \, \diff \bm{\sigma} 
  + 
  \int_E 
  \left( 
  \psi_j \, \bm{\nabla} \psi_i \, 
  \frac{\partial \tensor{F}}{\partial \mathbf U}  
  + \psi_i \, \psi_j\, 
  \frac{\partial \mathbf{S}}{\partial \mathbf{U}} \,
  \right) 
  \diff \bm{x} 
  . 
\label{jac1}
\end{align}
Similarly, the off-diagonal Jacobian can be obtained by taking the
derivative of \eqref{euler3} with respect to $\mathbf{u}_j$, in which
the host cell ``L'' is surrounded by the neighboring cells marked by ``R'':
\begin{equation}
  \frac{\partial \mathbf{R}_i}{\partial \mathbf{u}_j^{\rm R}}
  = 
  - \int_{\partial E} 
  \psi_i \, 
  \frac{\partial \widetilde{\tensor{F}}(\mathbf{U}_{\rm L},\, 
    \mathbf{U}_{\rm R})}
       {\partial \mathbf{U}_{\rm R}} 
  \frac{\partial \mathbf{U}_{\rm R}} 
       {\partial \mathbf{u}_j^{\rm R}}  
  \, \diff \bm{\sigma}
  = 
  - \int_{\partial E} 
  \psi_i^{\rm L} \, \psi_j^{\rm R} \, 
  \frac{\partial \widetilde{\tensor{F}}(\mathbf{U}_{\rm L},\, 
    \mathbf{U}_{\rm R})} 
       {\partial \mathbf{U}_{\rm R}}
  \, \diff \bm{\sigma}
  .
\label{jac2}
\end{equation}
The Riemann flux Jacobian matrices 
$\partial \widetilde{\tensor{F}}(\mathbf{U}_{\rm L},\, \mathbf{U}_{\rm
  R}) / \partial \mathbf{U}_{\rm L}$, $\partial \widetilde{\tensor{F}}
(\mathbf{U}_{\rm L},\, \mathbf{U}_{\rm R}) / \partial \mathbf{U}_{\rm
  R}$ in \eqref{jac1} and \eqref{jac2} are evaluated exactly through
 the automatic differentiation (AD), others can be derived
easily.

The global Jacobian matrix $\tensor{J}$ is made of the diagonal and
off-diagonal matrices above.  When ${\bm{\sigma}}$ is an interior
face, the flux $\widetilde{\tensor{F}}(\mathbf{U}_{\rm L},\,
\mathbf{U}_{\rm R})$ is calculated with Roe's Riemann solver
\cite{Roe1981jcp}.  When $\bm{\sigma}$ is a boundary face with a
appropriate boundary condition, one has
\begin{equation}
  \widetilde{\tensor{F}}_{\mbox{\scriptsize bc}} 
  = 
  \widetilde{\tensor{F}}(\mathbf{U}_{\rm L},\, \mathbf{U}_{\rm ghost}),
\end{equation}
where $\mathbf{U}_{\rm{ghost}}$ is a function of $\mathbf{U}_{\rm L}$
corresponding the boundary condition, and $\widetilde{\tensor{F}}$ is also
consistently computed by the same Roe's Riemann solver used on the
interior faces.  Then,  the boundary flux Jacobian matrix can be expressed as
\begin{equation}
  \frac{\partial \widetilde{\tensor{F}}_{\mbox{\scriptsize bc}} }
       {\partial \mathbf{U}_{\rm L}} 
  =
  \frac{\partial \widetilde{\tensor{F}}}{\partial \mathbf{U}_{\rm L}}
  +
  \frac{\partial \widetilde{\tensor{F}}}{\partial \mathbf{U}_{\rm
      ghost}} 
  \frac{\partial \mathbf{U}_{\rm ghost}}{\partial \mathbf{U}_{\rm L}}
    .
\end{equation}
The Jacobian matrix $\bm{\nabla} \psi_j \, {\partial \tensor{F}} /
{\partial \mathbf{U}}$ in the volume integration and the source-term
Jacobian matrix ${\partial \mathbf{S}} / {\partial \mathbf {U}}$ are
given by \eqref{dfdu} and \eqref{dsdu}, respectively, in the Appendix.
The Jacobians $\partial \widetilde{\tensor{F}} / \partial
\mathbf{U}_{\rm L}$, $\partial \widetilde{\tensor{F}} / \partial
\mathbf{U}_{\rm ghost}$ and $\partial \mathbf{U}_{\rm ghost} / \partial
\mathbf{U}_{\rm L}$ are obtained exactly by the automatic
differentiation (AD).

\section{Time-stepping strategy} 
\label{sec:timestepping}
 In this section, the time-stepping strategy of the eMG framework is discussed as a time-marching solver to compute the steady solutions of the Euler equations. 
There are two different time steps needed to be determined. One for the PRK time stepping $\delta \tau$ and the other for EXP1 smoothing which is empirically chosen as large as $(p_{\scriptsize {\mbox {max}}}+1) \, \delta \tau$. As such, only $\delta \tau$ should be determined.  
$\delta \tau$ is determined by
\begin{equation}
 \delta \tau 
 =   
 \frac{\mbox{CFL} \, h_{\rm 3D}}
      {\left(2 p+ 1\right) \left(\|\bm{v}\|+ c \right)} ,
 \quad
 h_{\rm 3D}
 :=  
 2 d 
 \frac{| E |}{| \partial E |},
\label{x3d}
\end{equation}
where $\mbox{CFL}$ is the global Courant-Friedrichs-Lewy
  (CFL) number,
$p$ the accuracy level,
$\bm{v}$ the velocity vector at the cell center, 
$c$ the speed of sound,
$d$ the spatial dimension,
$| E |$ and $| \partial E |$ are the volume and the surface area
of the boundary of $E$, respectively;
and $h_{\rm 3D}$ represents a characteristic size of a cell in 3D
defined by the ratio of its volume and surface area.
All the methods mentioned in this paper have been  implemented in the HA3D flow solver developed by the author, which is for solving three-dimensional problems
as its name indicates. So in order to support 2-D computations, 
a 2-D mesh is extruded to a 3-D (quasi-2D) mesh by one layer of cells and 
we use $h_{\rm 2D}$ instead of $h_{\rm 3d}$ to eliminate the effect of the $z$ dimension on obtaining the truly 2-D time step.  
Given the cell size $\Delta z$ in the $z$ direction, ${h}_{\rm 2D}$ is determined by
\begin{equation}
  \frac{2}{h_{\rm 2D}} 
  = 
  \frac{3}{{h}_{\rm 3D}} - \frac{1}{\Delta z}.
\label{x2d}
\end{equation}
To enhance the computational efficiency for the steady problems, the \mbox{CFL} number
  of both schemes are dynamically determined by the following formula
\begin{subequations}
\begin{align}
  &
  \mbox{CFL}_n
  = 
  \min 
  \left\{ 
  \mbox{CFL}_{\max}, \, 
  \max
  \left[ \| R(\rho_n) \|_2^{-1},\, 1
    + \frac{( n - 1 )}{ (2p+1) } 
    \right] 
  \right\} 
  ,
\label{cfl}
  \\
  &
  \| R(\rho_n) \|_2 
  := 
  \frac{1}{|\Omega|}
  \left[ {\displaystyle \int_\Omega R(\rho_n)^2 
      \diff \bm{x}  } \right]^{1/2}
  ,
\end{align}
\end{subequations}
 where $R(\rho_n)$ denotes the residual of density,
$\mbox{CFL}_{\max}$ is the user-defined maximal CFL number, $n$ is the
number of iterations, and $p$ is the spatial order of accuracy.  Thus,
such a CFL evolution strategy produces a robust start up when the initial flow flied is in a strong nonlinear evolvement 
and increases the time-step size exponentially later to improve computational efficiency.
In all the test cases considered, the upper-bound CFL number of \eqref{cfl} is taken as follows: $\CFLmax=10^3$ for the implicit BE method; $\CFLmax=10^2$ for the eMG method.

\section{Numerical Results} 
\label{sec:results}

This section presents the results of two typical steady flow cases:
flows past a circular cylinder in qusai-2D and a sphere in 3D at Mach number \Ma=0.3. 
The results are computed by the new V-cycle eMG method,
and are compared with the results obtained by a fast fully implicit scheme:  
the first-order backward Euler (BE) method solved by the ILU preconditioned GMRES method.
The same parameter setting of Krylov subspace is used for both the exponential and the implicit methods, 
where the dimension of the Krylov basis $m$ is 30 and the convergence tolerance of the Krylov subspace is  $10^{-5}$.

\subsection{Flow over a circular cylinder in quasi-2D}
\begin{figure}[htbp!]
\centering 
\includegraphics[width=0.618\columnwidth]{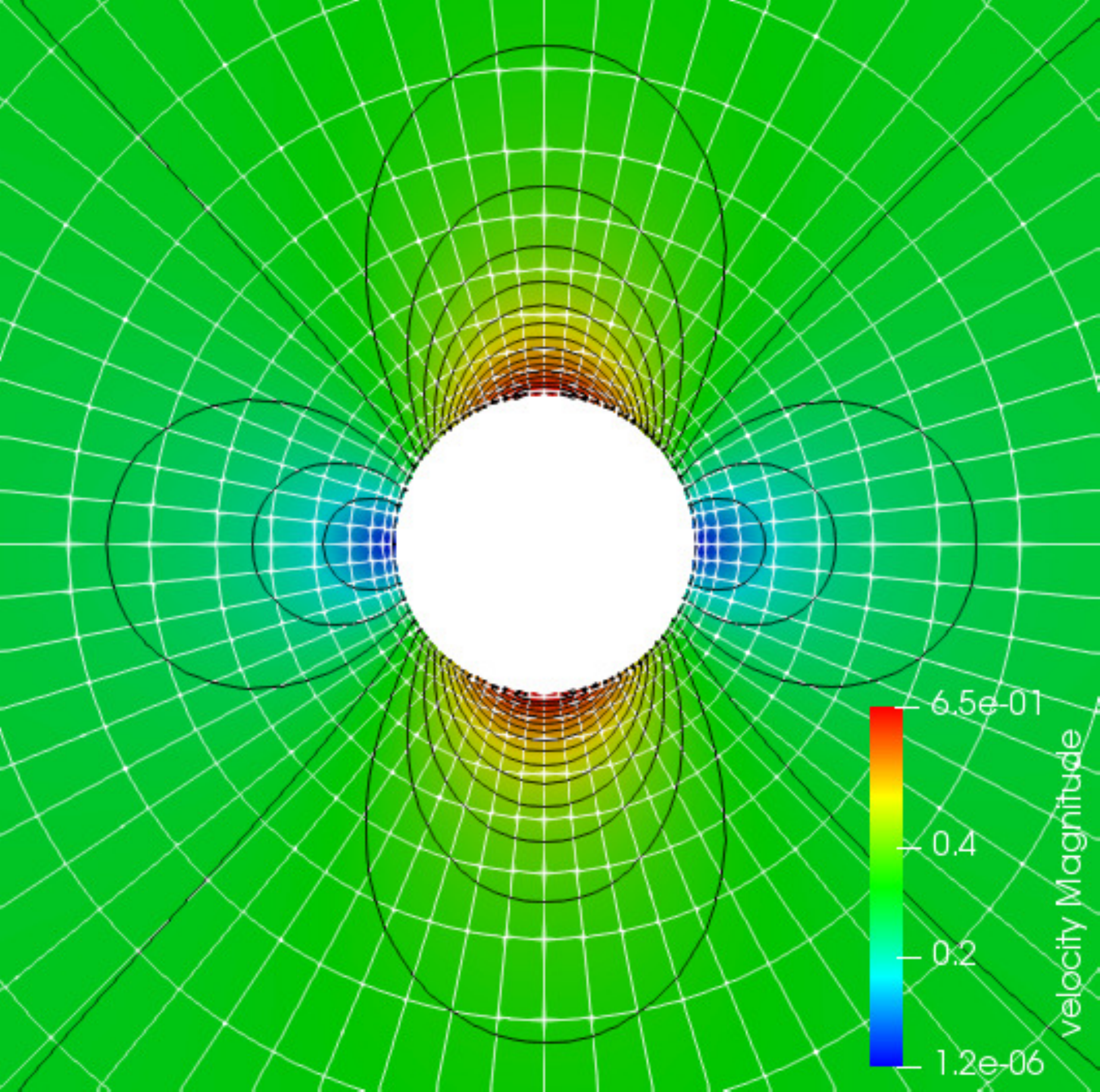} 
\caption{Flow contour computed for the flow past a cylinder at $\Ma=0.3$ with eMG and DG $p=3$}
\label{grid_cylinder}
\end{figure}

In this case, the results obtained for a flow over a circular cylinder at Mach number $\Ma = 0.3$ is presented.
The cylinder has a radius of 1 and surrounded in a circular computational domain of radius $5$, as shown
in Fig.~\ref{grid_cylinder}.  The quasi-2D mesh contains 896 quadratic curved hexahedral elements. 
The inviscid solid wall boundary condition is imposed on the inner wall surface and the out circular boundary is set as the far-field characteristic boundary condition.

\begin{figure}[htbp!]
\centering 
\includegraphics[width=0.75\columnwidth]{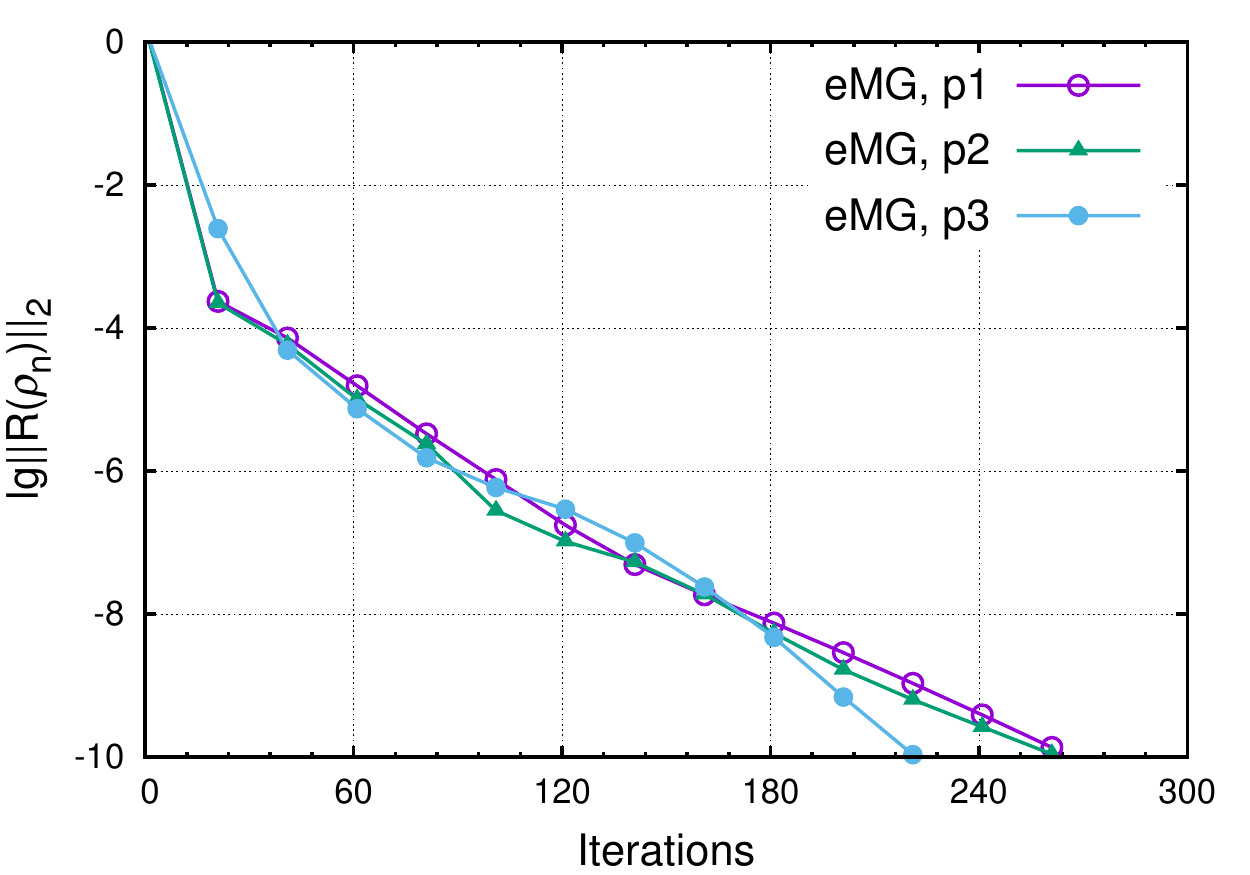} 
\caption{$p$-independent convergences with the eMG method} 
\label{2d_pmg}
\end{figure}

\begin{figure}[htbp!]
\centering 
\includegraphics[width=0.75\columnwidth]{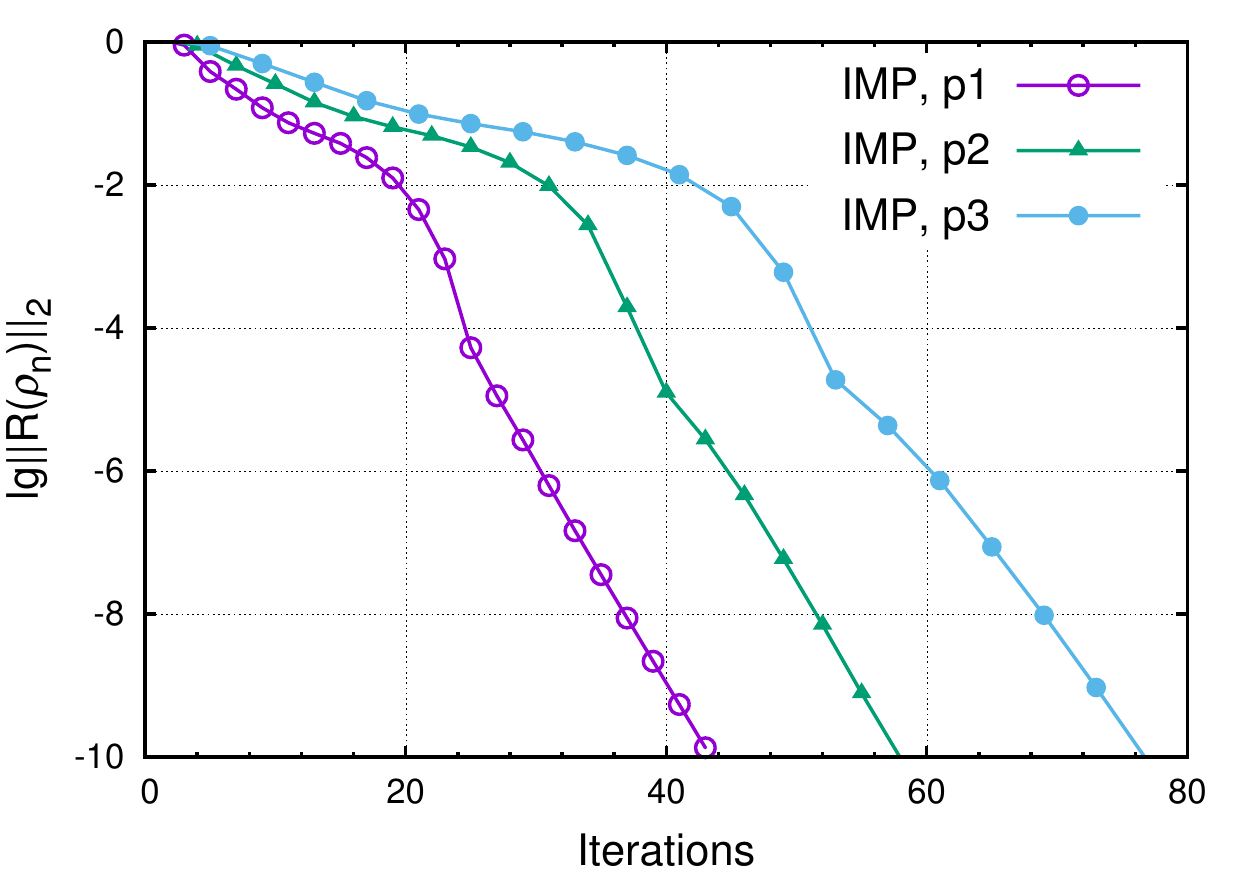}
\caption{Convergence histories of the implicit method with varying spatial accuracy} 
\label{2d_imp}
\end{figure}

\begin{figure}[htbp!]
\centering 
\includegraphics[width=0.75\columnwidth]{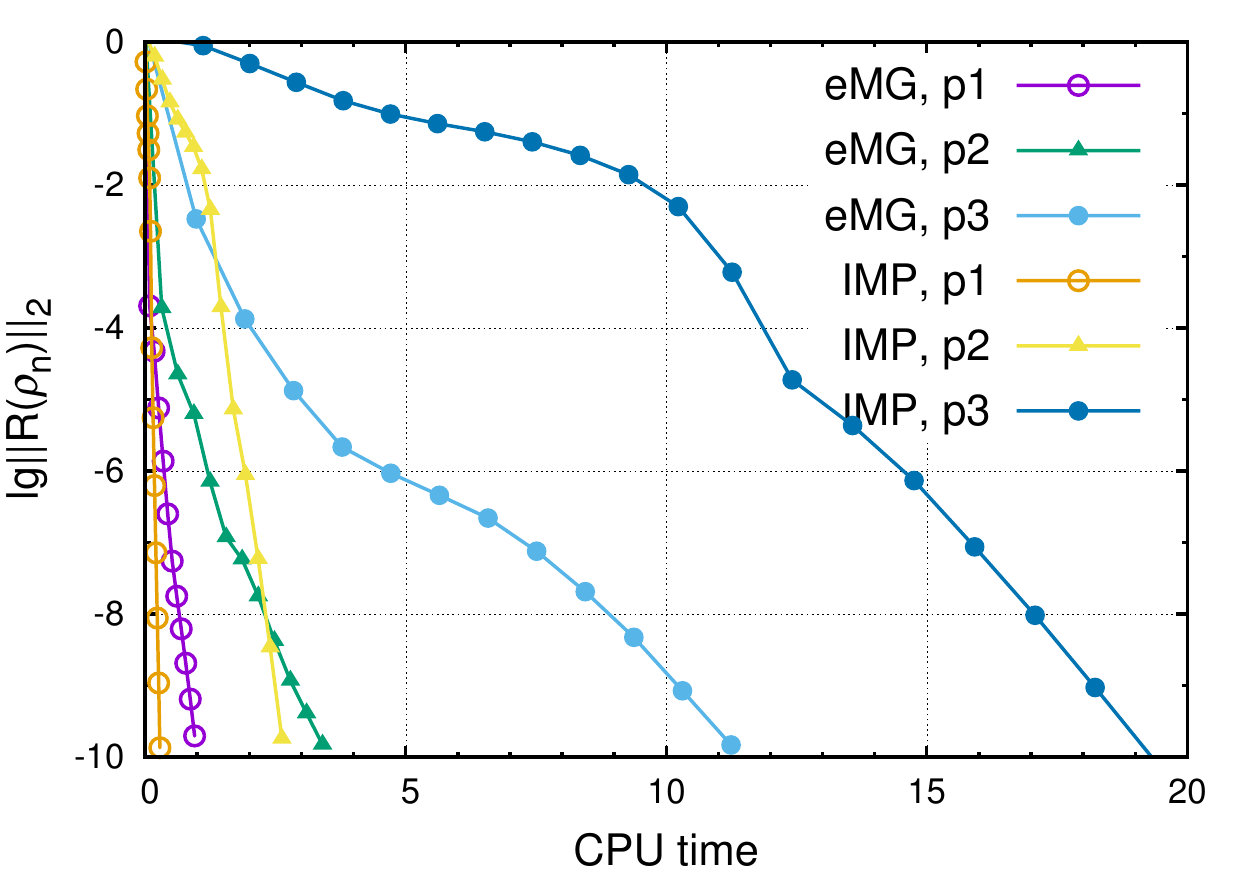}
\caption{Performance comparison between the eMG method and the implicit method at different spatial accuracy}
\label{com2d}
\end{figure}

In Fig.~\ref{2d_pmg}, the $L_2$ norm of density
residual $R(\rho_n)$ is plotted versus the iteration by using the eMG scheme, 
indicating convergence rates independent of spatial order of accuracy $p$, or say $p$-independent.
The results obtained with a fast, implicit ILU preconditioned GMRES is computed in Fig.~\ref{2d_imp}, 
which shows the convergence histories of the implicit method with varying spatial accuracy. 
The results show rapid quadratic Newton convergences which are actually dependent on the spatial order of accuracy $p$. 
To see how promising is the eMG performance comparing with the fully implicit method, 
the two results are compared in Fig.~\ref{com2d}, where the CPU time is normalized by that of the eMG scheme. 
As we can see, the implicit method (IMP) is faster for $p=1,2$ cases, but is slower than the eMG scheme for the $p=3$ case. 
So for high-order computations, the eMG method is at least comparable to the implicit method in term of overall performance.

\subsection{Flow over a sphere in 3D}

\begin{figure}[htbp!]
\centering 
\includegraphics[width=0.618\columnwidth]{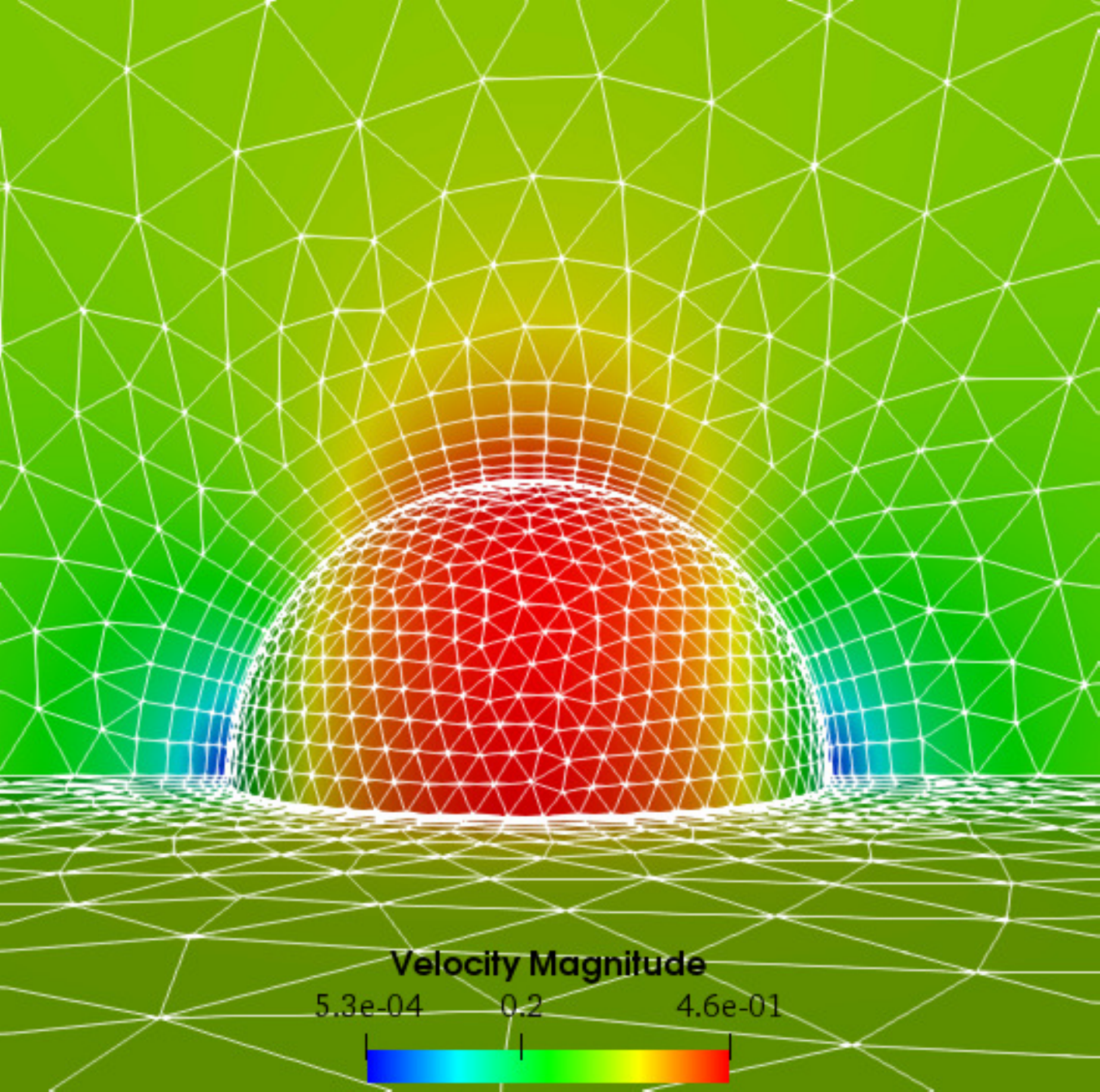}%
\caption{Flow contour computed for the flow past a sphere at $\Ma=0.3$ with eMG and DG $p=3$}
\label{3d_contour}
\end{figure}

The computational efficiency of the
eMG scheme is investigated for a three-dimensional flow past a
sphere with the Mach number $\Ma = 0.3$, representing a basic benchmark of 3-D flow problems. 
The radius of the sphere is 1 and the radius of far-filed spherical shell is 5.
The sphere surface is set as a slip wall boundary condition, 
and the outer boundary uses a far-field characteristic boundary condition with Riemann invariants.
The mesh respects the flow symmetries of the horizontal and vertical
planes, on which a symmetry boundary condition is imposed.
The generated curved mesh consists of 9778 tetrahedrons and 4248 prisms, 14026 cells in total.
A close-up view of the mesh about the sphere and the velocity contour
computed with the eMG scheme at $p=3$ is illustrated
in Fig.~\ref{3d_contour}.

\begin{figure}[htbp!]
\centering 
\includegraphics[width=0.75\columnwidth]{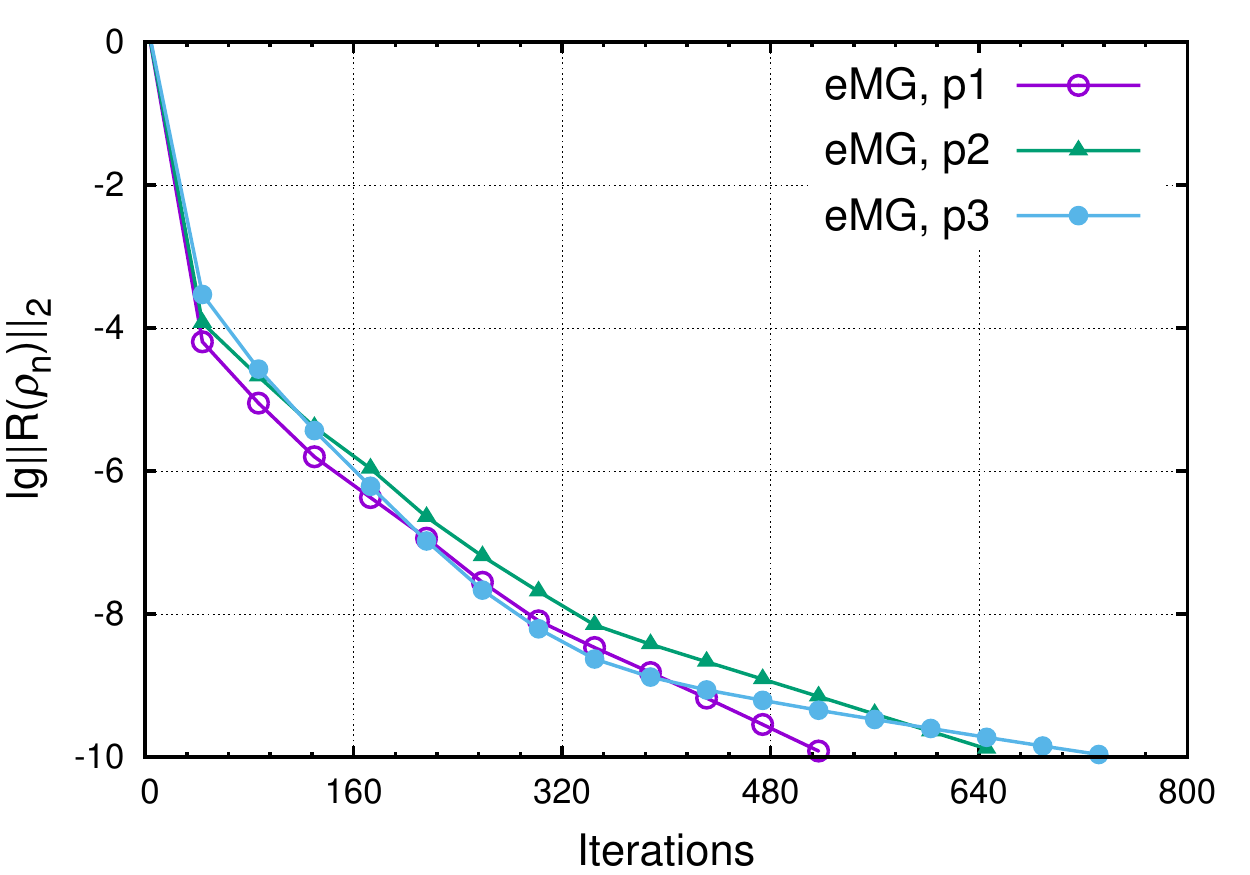} 
\caption{$p$-independent convergences with the eMG method} 
\label{3d_pmg}
\end{figure}

\begin{figure}[htbp!]
\centering 
\includegraphics[width=0.75\columnwidth]{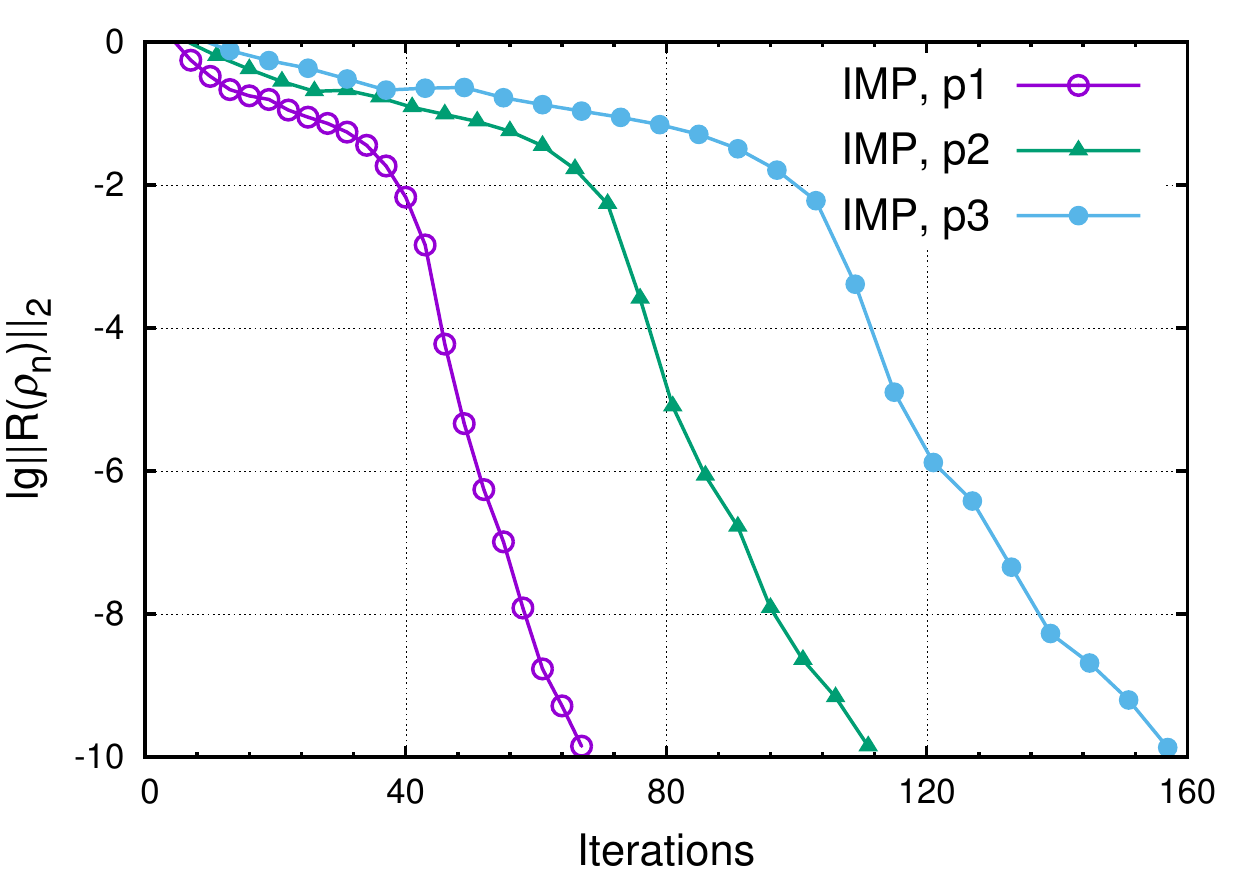}
\caption{Convergence histories of the implicit method with varying spatial accuracy} 
\label{3d_imp}
\end{figure}

\begin{figure}[htbp!]
\centering 
\includegraphics[width=0.75\columnwidth]{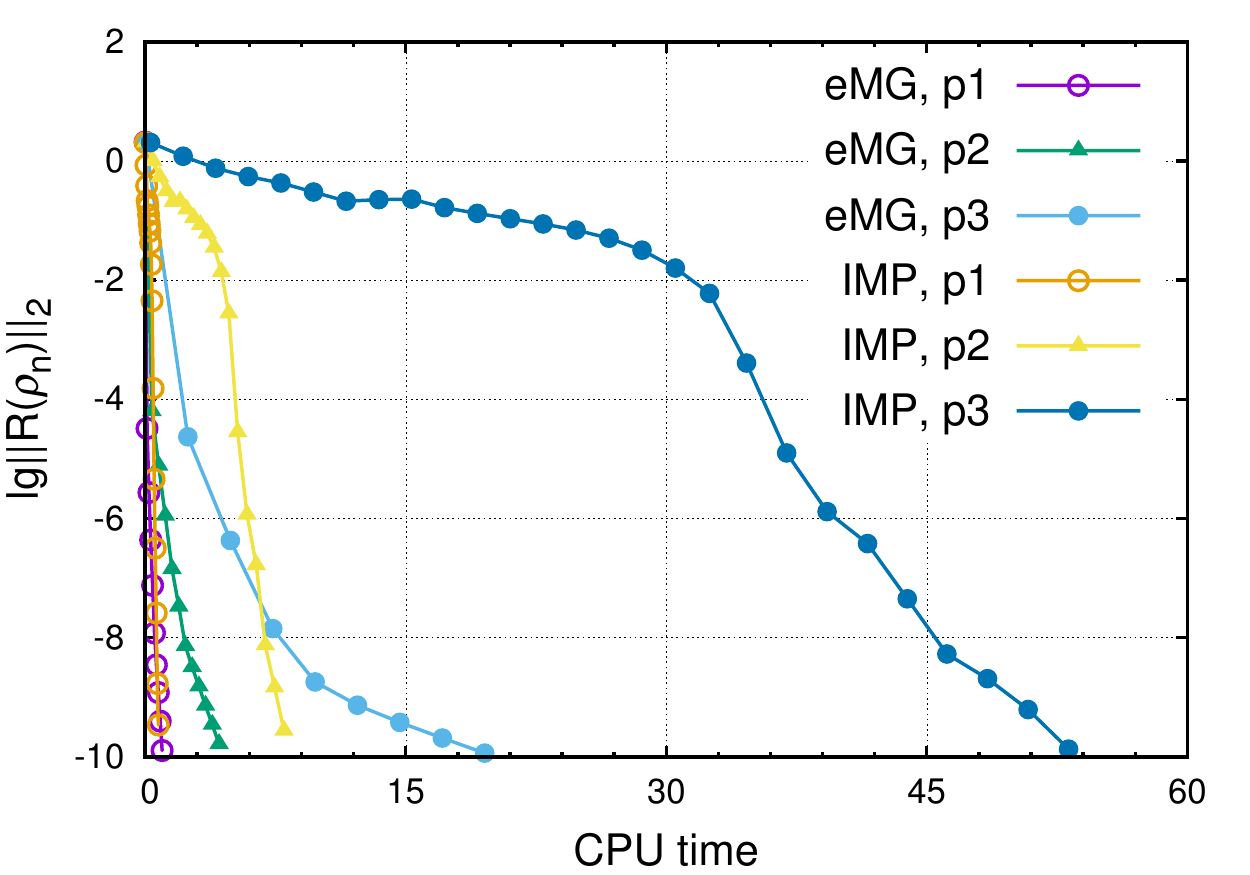}
\vskip -0.5cm
\caption{Performance comparison between the eMG method and the implicit method at different spatial accuracy}
\label{3d_vs}
\end{figure}

Fig.~\ref{3d_pmg} shows the convergence histories of the
 eMG method for spatial order of accuracy $p=1 \sim 3$. 
Again, $p$-independent convergences do appear. 
In Fig.~\ref{3d_imp}, convergence histories of the implicit method (IMP) are shown with iteration counts.
Fig.~\ref{3d_vs} compares both methods measured in CPU time. As one can see that although IMP is fast
in terms of iteration counts, the computational cost per iteration is relatively high and the resulting CPU time is penalized. 
Actually, when using high-order spatial schemes along with an implicit method, 
the high-order global Jacobian matrix also consumes a large mount of memory.
The most significant part of memory usage ($\mbox{M}$) of the two methods eMG and IMP are compared as follows 
\begin{equation}
  \mbox{M}_{\mbox{\scriptsize{emg}}}
  = \mbox{NE} \left[\frac{5}{3}(p+1)(p+2)(p+3) + 150 \right];\\
  \quad
  \mbox{M}_{\mbox{\scriptsize{imp}}} 
  = 6\mbox{NE}\left[\frac{5}{6}(p+1)(p+2)(p+3) \right]^2
\end{equation}
For problems sized up to $\mbox{NE}=10^5$ elements at $p=3$, fourth-order spatial accuracy, a fully implicit method requires 45GB memory only
for storing the Jacobian matrix, while eMG only requires 0.03GB memory for storing the solution vectors plus the first-order Jacobian matrix for the same sized problem. Therefore, the eMG method is far more memory friendly compared with a fully implicit method, providing a more practical while efficient strategy  for solving steady problems with high-order methods.


\section{Conclusions}
\label{sec:final}
 The first-order exponential time integration scheme, EXP1, has been applied
to the $p$-multigrid DG framework combining with the PRK method. 
The algorithms and the physical natures of the methods are discussed.
The performance and memory usage are investigated and compared with the
fully implicit method solved with the ILU-GMRES linear solver.
Both 2-D and 3-D problems are computed to demonstrate the effectiveness of using
eMG method for the computations of steady flows. All the results exhibit 
$p$-independent convergence rates as expected for $p=1\sim 3$ order of accuracy.
Besides the memory friendly feature, the eMG scheme inherits the strong damping nature of the EXP1 scheme 
as shown in our previous works \cite{Shujie17AIAA, Shujie18AIAA,Shujie18JCP}.
It is observed that the eMG scheme uses shorter CPU time for $p=3$ cases and more efficient 
for the 3-D case compared to the fully implicit method. 

In conclusion, the exponential time integration method has been extended to 
the V-cycle $p$-multigrid DG framework for efficiently solving steady flows. Comparing to the fully implicit method, the eMG framework is much more memory friendly while achieving comparable computational efficiency, providing a new viable methodology for practical 3-D steady problems especially for high-order spatial discretizations.

\section*{Acknowledgments}

This work is funded by the National Natural Science Foundation of
China (NSFC) under the Grant U1530401.  The computational resources
are provided by Beijing Computational Science Research Center
(CSRC). SJL would like to thank Prof.Z.J.~Wang of University of Kansas
for the discussions on the time-stepping methods.

\appendix

\section{The Jacobian matrices}

The matrix $\bm{\nabla} \psi \, \partial \tensor{F} /
\partial \mathbf{U}$ in \eqref{jac1} is
%
\begin{equation}       
\left(                 
\begin{array}{ccccc}       
  -B_2 & \psi_x & \psi_y & \psi_z & 0
  \\ 
  a_0 \psi_x - B_1 u &  B_1 - B_2 - a_3 u \psi_x & u \psi_y - a_2 v
  \psi_x &  u \psi_z - a_2 w \psi_x & a_2 \psi_x
  \\  
  a_0 \psi_y - B_1 v &  v \psi_x - a_2 u \psi_y &  B_1 - B_2 - a_3 v
  \psi_y &  v \psi_z - a_2 w \psi_y & a_2 \psi_y
  \\  
  a_0 \psi_z - B_1 w & w \psi_x - a_2 u \psi_z & w \psi_y - a_2 v
  \psi_z &  B_1 -B_2  - a_3 w \psi_z & a_2 \psi_z
  \\  
  (a_0-a_1)  B_1 & a_1 \psi_x - a_2 B_1 u & a_1 \psi_y - a_2 v  B_1 &
  a_1 \psi_z - a_2 B_1 w & \gamma B_1 - B_2 \\
  \end{array}
\right),
\label{dfdu}
\end{equation}
where $\bm{v} := (u,\, v,\, w)$, $\bm{\omega} := (\omega_x,\,
\omega_y,\, \omega_z)$, $ \bm{\nabla} \psi
:= (\psi_x,\, \psi_y,\,  \psi_z)$, 
%
\begin{equation}
\begin{aligned}
  &
  a_0 
  := 
  \frac{1}{2} (\gamma -1) (u^2 + v^2 + w^2) 
  ,
  \quad
  a_1 
  := 
  \gamma e - a_0 ,
  \quad
  a_2 
  := 
  \gamma - 1,
  \quad
  a_3 
  := 
  \gamma - 2, 
  \\
  &
  B_1 
  := 
  \bm{v} \cdot \bm{\nabla} \psi
  =
  u \psi_x  + v \psi_y  + w \psi_z ,
  \quad
 B_2
  :=
  (\bm{\omega} \times \bm{x} ) \cdot \bm{\nabla}{\psi} .
\end{aligned}
\end{equation}
%

The source-term Jacobian matrix $\partial \mathbf{S} /\partial
\mathbf{U}$ in \eqref{jac1} is 
%
\begin{equation}      
  \frac{\partial \mathbf{S}}{\partial \mathbf{U}} =
  \left(                 
  \begin{array}{ccccc}       
    0& 0& 0& 0& 0\\ 
    0 & 0& -\omega_z & \omega_y & 0\\  
    0 & \omega_z & 0 & -\omega_x & 0 \\
    0 & -\omega_y & \omega_x & 0 & 0 \\
    0 & 0 &0 &0 & 0 \\
  \end{array}
  \right)
  .
\label{dsdu}   
\end{equation}
%





\end{document}